\documentclass[%
 reprint,
 amsmath,amssymb,
 aps,
]{revtex4-2}

\usepackage{graphicx}% Include figure files
\usepackage{dcolumn}% Align table columns on decimal point
\usepackage{bm}% bold math
\usepackage[dvipsnames]{xcolor}
\begin{document}

\preprint{Preprint}

\title{Testing the Weak Equivalence Principle with Cosmological Gamma Ray Bursts}

\author{Matipon Tangmatitham}
\email{matipon@gmail.com}
\affiliation{Department of Physics, Michigan Technological University, 1400 Townsend Drive, Houghton, MI  49931 USA}

\author{Robert J. Nemiroff}
 \email{nemiroff@mtu.edu}
\affiliation{Department of Physics, Michigan Technological University, 1400 Townsend Drive, Houghton, MI  49931 USA}

\date{\today}% It is always \today, today,

\begin{abstract}
Gamma Ray Bursts (GRBs) with rapid variations at cosmological distances are used to place new limits on violations of the gravitational weak equivalence principle (WEP). These limits track intrinsic timing deviations between GRB photons of different energies as they cross the universe, in particular in the KeV to GeV energy range. Previous limits in this energy range have involved only the gravitational potential of local sources and utilized temporal variability on the order of 0.1 seconds. Here WEP violation limits are derived from sources with greater distance, faster variability, and larger intervening mass. Specifically, GRB sources with redshifts as high as 6.5 are considered, with variability as fast 0.2 milliseconds, and passing the gravitational potentials of inferred clusters of galaxies distributed randomly around the line of sight. WEP violation limits are derived from data from GRB 910711, GRB 920229, GRB 021206, GRB 051221, GRB 090429, and GRB 090510. The strongest constraint in the very early universe comes from GRB 090429 which limits $\gamma(500$ keV$) - \gamma(250$ keV$) < 1.2\times10^{-13}$. The strongest overall constraint comes from GRB 090510 which yields a WEP violation limit of $\gamma(30$ GeV$) - \gamma (1$ GeV$) < 6.6\times10^{-16}$. This strongest constraint is not only a new record for WEP violation limit for gamma-ray photons and in the early universe, but the strongest upper bound for $\Delta \gamma$ that has ever been recorded between any two energy bands. 
\end{abstract}

\maketitle

\section{Introduction}

The gravitational equivalence principle has been discussed for over 300 years. Newton's statement of this principle was that mass and weight are locally measured to have an identical ratio for all bodies \cite{1687pnpm.book.....N}. Einstein once stated the equivalence principle as ``the acceleration imparted to a body by a gravitational field is independent of the nature of the body" \cite{2004mere.book.....E}. A special case of this, the Weak Equivalence Principle (WEP), applies only to freely falling objects that are not themselves gravitationally bound. In other words, any two free-falling test particles must follow the same trajectory.

In this work, the WEP is explored with photons from gamma-ray bursts (GRBs) of very high energy, rapid variability, and cosmological distances. On their way to the Solar System, each GRB photon must travel through localized gravitational fields of galaxies and clusters of galaxies. Each field induces a temporal lag known as the Shapiro time delay \cite{1964PhRvL..13..789S}. Parameterized post-Newtonian (PPN) deviations from general relativity (GR) can be described by a factor $\gamma$ where $\gamma = 1$ correspond to standard GR \cite{Will2006}. The Shapiro time delay for a photon can be expressed as
\begin{equation}\label{eqn:shapiro}
    t_{Shapiro} = -\frac{1 + \gamma}{c^3} \int_{D_S}^{D_O} U(\mathbf{r}(t),t) dr ,
\end{equation}
where the gravitational potential $U$ is integrated along radial coordinate $r$ from a source at light-travel distance $D_S$ to the observer at light-travel distance $D_O$. For each intermediate mass near the light path of mass $M$, assumed small compared to $D_S$, this time delay in the frame of the intermediate mass can be described as
\begin{equation}\label{eqn:shapiroapprox}
    t_{Shapiro} \approx (1 + \gamma) \frac{G M}{c^3} \ln \frac{4 D_M D_{MS} }{b^2} ,
\end{equation}
where $b$ is the comoving impact parameter of the light with respect to the center of the intervening mass near the light path, $D_M$ is the light-travel distance to the point of closest approach to the intervening mass, $D_{MS}$ is the light-travel distance between $D_M$ and $D_S$.

WEP violations can be parameterized by photons of different energies having nonzero $\Delta \gamma$, resulting from a nonzero $\Delta t_{Shapiro}$. A small change in $\Delta \gamma$ in the limit where deviation from GR is small ($\gamma \approx 1$), can be characterize as $\Delta t_{Shapiro} = \Delta \gamma \  t_{Shapiro}/2$.

When photons released from a source arrive at a detector at different times, the time difference observed, $\Delta t_{Obs}$, is a combination of the intrinsic time of release from the source $\Delta t_{Release}$ and the difference in Shapiro time delays experienced by the two photons $\Delta t_{Shapiro}$. When observed today, $\Delta t_{Obs} = (1+z_S) \Delta t_{Release} + (1+z_M) \Delta t_{Shapiro}$, where each of the time differences is further expanded by the scale factor $1/a = 1+z$ expressed in terms of redshifts of the source $z_S$ and the intermediate mass $z_M$.

Even though there is no way to directly measure the $\Delta t_{Release}$ from a source, we can attribute the upper limit of WEP violations to $\Delta t_{Shapiro} (1 + z_M) \lessapprox \Delta t_{Obs}$. Although a nonzero $(1+z_M) \Delta t_{Shapiro}$ could theoretically cancel out with $(1+z_S) \Delta t_{Release}$, it would be practically impossible to have it cancel out for every pair of photons having different $\Delta \gamma$. Therefore, by observing $\Delta t_{Obs}$ we can put an upper limit to $\Delta \gamma$ of the corresponding energies such that $\Delta \gamma \lessapprox 2 \Delta t_{Obs}/[(1+z_M) t_{Shapiro}]$.

WEP violation limits have been set previously using this method on non-GRB variables. A earlier notable limit on $\Delta \gamma$ in a cosmological setting is from a fast radio burst (FRB) where $\gamma(1.23 \text{GHz})$ $-$ $\gamma(1.45 \text{GHz}) < 4.36 \times 10^{-9}$ \cite{2015PhRvL.115z1101W}. Similarly, a limit on $\Delta \gamma$ between gravitational wave (GW) and associated electromagnetic signal could put the limit in the order of $\Delta \gamma < 10^{-10}$ \cite{2016PhRvD..94b4061W} and, with time delay of less than 1.7 s in GW 170817, has been measured to be $\gamma_{GW} - \gamma_{EM} < 9.8 \times 10^{-8}$ \cite{2018PhRvD..97d1501B}. 

WEP violation limits involving only GRBs have also been published. Precedents include an analysis of GRB 090510 by Gao et al. \cite{Gao}, resulting in $\gamma_{\text{GeV}} - \gamma_{\text{MeV}} < 2 \times 10^{-8}$. Nusser \cite{Nusser} further computed a WEP violation limit from GRB 090510 to be $\gamma_{\text{GeV}} - \gamma_{\text{MeV}} < 2.3 \times 10^{-12}$. Both of these used a $\Delta t_{Obs}$ on the order of 0.1 seconds.

All of these previous limits on WEP violations, however, have only taken into account the gravitational potential of our Milky Way galaxy. One practical reason is that the gravitational potential of the Milky Way is known, whereas other potentials along each photon path are not known. In this work, gravitational sources that must exist, statistically, near a random light path to a cosmologically distant object are generated. From a collection of these random light paths, a distribution of expected Shapiro time delays are generated. It will be demonstrated that even the shortest Shapiro time delays expected are, statistically, much larger than the Shapiro time delays created by the Milky Way galaxy or any other local potential. 

\section{Method}

\begin{figure}
\centering
\includegraphics[width=0.45\textwidth]{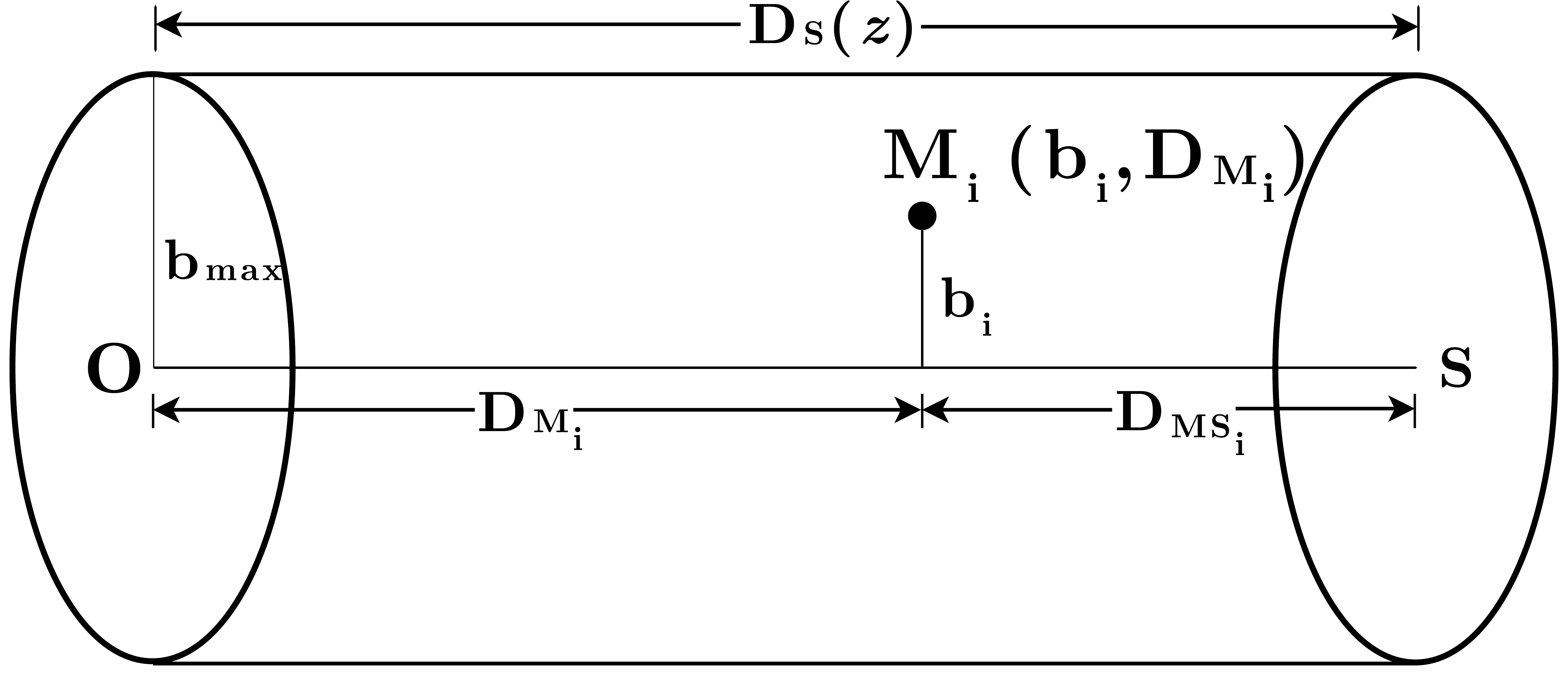}
\caption{\label{fig:diagram}For a source located at $D_S(z)$, each intermediate mass ($M_i$) near the light path between the observer ($O$) and the source ($S$) is placed randomly at comoving cylindrical coordinate $(r,z) = (b_i,{D_M}_i)$.}
\end{figure}

In this work, a distribution of Shapiro time delays will be simulated from a uniformly random distribution of clusters of galaxies near the light path to a source at redshift $z_S$. A standard concordance cosmology is used ($\Omega=1$, $H_0 = 67.74$ km sec$^{-1}$ Mpc$^{-1}$, $\Omega_\Lambda = 0.6911$, $\Omega_m = 0.3089$)\cite{2016A&A...594A..13P}. For a source located at comoving distance $D_{S}(z)$, the light path to the observer will pass near a series of galaxy clusters, each of mass $M_i$, at comoving distance ${D_M}_i$ and comoving impact parameters $b_i$ (Fig. \ref{fig:diagram}). 

For $n$ intermediate clusters, the total Shapiro time delay can be expressed as 
\begin{equation}\label{eqn:shapirosum}
    t_{Shapiro} = (1+\gamma)\sum_{i=1}^n (1+{z_M}_i)\frac{G M_i}{c^3} \ln \frac{4{D_M}_i {D_{MS}}_i}{{b_i}^2} .
\end{equation}
Each mass $M_i$ is placed at comoving cylindrical coordinate $(r,z) = (b_i,{D_M}_i)$ with $b_i \in \left[0,b_{max}\right]$ and ${D_M}_i \in \left[0,D_S(z)\right]$. The cluster distribution within the comoving cylinder is assumed to be randomly distributed with uniform density $\rho_{crit} \Omega_C$. A lower limit of $\Omega_C = 0.15$, taken from Bahcall et al.  \cite{2003ApJ...585..182B}, is the fraction of mass attributed to clusters of galaxies. 

Each cluster is assumed to have mass between $10^{12} - 10^{15} M_\odot$ and the frequency distribution described by the mass function of clusters of galaxies\cite{1993ApJ...407L..49B}  
\begin{equation}\label{eqn:massfunctionofclusters}
    n(>M) = 4\times10^{-5}\left(\frac{M}{M^*}\right)^{-1}e^{-M/M^*} \text{h}^3 \text{Mpc}^{-3} ,
\end{equation}
where $M^* = 1.8\times10^{14} h^{-1} M_{\odot}$. 

The maximum impact parameter, $b_{max}$, and the comoving distance, $D_S(z)$, determine the volume and the amount of mass contained within the cylinder. As $b_{max}$ increases, clusters that are farther from light path are included, adding smaller contributions per cluster but collectively larger contributions to the Shapiro time delay. A characteristic $b_{max} = 10$ Mpc is chosen based on the impact parameters of known gravitational lensing effects from galaxies and clusters on background galaxies \cite{2013MNRAS.432.1544M, 2017MNRAS.467.3024L}. Clusters located within this cutoff radius are known to exhibit weak lensing distortions and therefore must also contribute to the Shapiro time delay. 

Although it is impossible to know the exact gravitational field that an observed pair of photons has passed through, a distribution of expected time delays for photons traveling from redshift $z_S$ can be bounded from a simulation to high accuracy. If $t_5$ represents the 5th percentile of Shapiro time delay generated by the simulation, then for over 95\% of random light paths, $t_5 < t_{Shapiro}$. Based on the value of $t_5$, we can calculate the upper limit for $\Delta \gamma < 2\frac{\Delta t_{Obs}}{t_5}$ with 95\% confidence interval, where $\Delta t_{Obs}$ is the measured minimum variability time scale between photons of different energies for specific GRBs.

\begin{figure}
\centering
\includegraphics[width=0.5\textwidth]{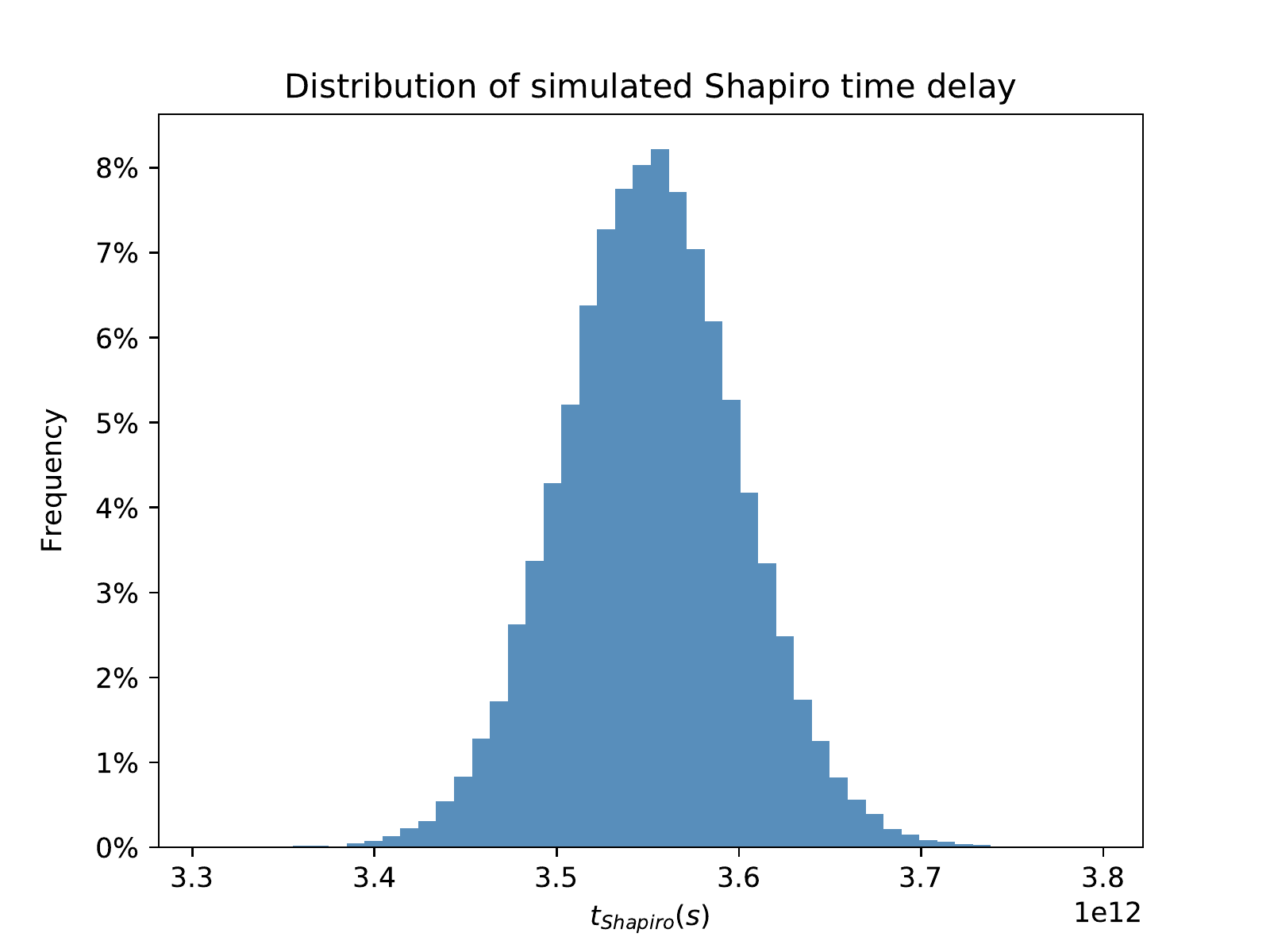}
\caption{\label{fig:histogram}Distribution of Shapiro time delay for a source located at $z_S=1$ and all clusters within comoving radius of $b_{max} = 10$ Mpc are considered.}
\end{figure}

Fig. \ref{fig:histogram} shows a histogram generated from adding the Shapiro time delays from uniformly distributed clusters with the mass distribution given by Eq. \ref{eqn:massfunctionofclusters} up to redshift $z_S = 1$, and $b_{max} = 10$ Mpc. Filling this cylindrical volume up to $\rho_{crit} \Omega_C$ is equivalent to adding $4.10 \times 10^{16} M_\odot$ to the volume. The resulting Shapiro time delay has average of $3.55 \times 10^{12}$ s. Among the distribution of Shaprio time delays generated, 95\% has $t_{Shapiro} > 3.47\times10^{12}$ s. If a detection threshold puts a limit $\Delta t_{Obs} < 1$ ms from an observation, this corresponds to $\Delta \gamma < 5.8\times10^{-16}$ with 95\% confidence interval. If we compare this to the time delay contribution from the Milky Way Galaxy with $M_{MW} = 6\times 10^{11}M_\odot$, $d = 1500$ Mpc, and $b = 5$ kpc, \cite{Nusser} this would only give $t_{Shapiro} = 7.0 \times 10^7$ s, almost five orders of magnitude smaller.

\section{Results}

While this method could limit WEP violation signals between any two particles released simultaneously from the same source, in this work the method is applied to GRBs expected to put the smallest upper bound on WEP violations. These will typically occur for GRBs with the most rapid intrinsic variabilities and the greatest distances. A list of chosen notable GRBs, their reported data, and corresponding limits on WEP-violation limits are shown in Table \ref{tab:deltagammatable}. Each GRB has its own details and caveats as described below. 

GRB 910711 was measured by BATSE \cite{1989BAAS...21..860F} and reported to have a total duration of 16 ms that persists between energy channel 2 and 4, with energy ranges from 100 KeV to 300 KeV \cite{1992Natur.359..217B}. Unfortunately this GRB did not have a measured redshift and so does not give direct evidence of coherence to the WEP. However, GRB 910711 is a triggered short GRB and so likely occurred at a redshift of $z > 0.1$, as have all triggered short GRBs to date  \cite{2014ARA&A..52...43B, GreinerTable}. 

GRB 920229 was detected by BATSE and reported to have a total duration of 190 ms but an internal flare with a rise time of 0.22 $\pm$ 0.03 ms \cite{1999ApJ...511L..89S, 1999PhRvL..82.4964S}. Inspection of Fig 2. of \cite{1999ApJ...511L..89S} indicates this rise was seen in three energy channels simultaneously with energies 25 - 50 keV, 50 -100 keV, and 100 - 300 keV respectively. Although photons likely came in across this entire energy range, the energy range for this rise time as quoted by \cite{1999PhRvL..82.4964S} was between 0.03 and 0.20 MeV. GRB 920229, a triggered short GRB, did not have a measured redshift, but as a lower limit, it can be assign a redshift below that of any triggered short GRB to date: $z > 0.1$. 

GRB 021206 was detected by RHESSI and reported to have a rapid flare of gamma-rays with a duration of about 15 ms between 2 MeV and ``above 10 MeV" \cite{2004ApJ...611L..77B}. Inspection of Fig. 2 in \cite{2004ApJ...611L..77B} indicates a conservative energy range estimate of between 3 and 10 MeV. Additionally, GRB 021206 had a computed pseudo-redshift of 0.3, listed as accurate to within a factor of two \cite{2004ApJ...611L..77B}. 

GRB 051221A was detected by gamma-ray detectors aboard both Konus-Wind and Swift. Analysis of the first three peaks in Konus-Wind data indicate that they occur within 4 ms of each other, in all three energy bands analyzed \cite{2006JCAP...05..017R}. Conservatively, the smallest energy gap would be between 0.07 and 0.3 MeV. This GRB had a spectroscopic redshift measured of 0.547 \cite{2006ApJ...650..261S}. 

GRB 090429B is chosen not for its rapid variability but for its high estimated (mean) redshift of 9.4. GRB 090429B was detected by the gamma-ray detector on Swift and contained three temporal peaks with a combined measured duration of 5 sec \cite{2009GCN..9281....1U}. Assuming a Small Magellanic Cloud dust law yields a photometric redshift bounds (90 \% confidence level) of $9.06 < z < 9.52$ \cite{2011ApJ...736....7C}. However, the lowest photometric redshift estimated from a high-z dust law gives $z > 6.5$ at 99 \% confidence. To be conservative, this lowest redshift estimate is used here. The measured $t_{90}$ duration of the burst was 5.2 sec \cite{2011ApJ...736....7C}, however, a temporal lag in the cross-correlation between an energy channel of 15 - 25 keV and an energy channel of 50 - 100 keV was found to be 1200 ms (95 \% confidence level) \cite{2011ApJ...736....7C}. 

GRB 090510 was detected by the Fermi satellite including its Large Area Telescope. Subsequent analysis resulted in a reported variability of 1.55 ms over an energy range from 1.58 GeV to about 24.7 GeV \cite{2012PhRvL.108w1103N}. The 2-$\sigma$ lower limit on this GRB's spectroscopic redshift is 0.897 \cite{2009Natur.462..331A}. 

These WEP-violation limits assume that the mass distribution of clusters of galaxies given by  (\ref{eqn:massfunctionofclusters}) have been relatively unchanged from $z=6.5$ to today. This simulation also assumed that the Universe is homogeneous in the comoving coordinate across the volume within the cutoff radius of $b_{max} = 10$ Mpc.

\begin{table*}[t]
\begin{center}
\begin{tabular}{| l | c | c | c | c | c | c |}
\hline
Name      &    Instrument     & $\Delta t_{Obs}$     &  $E_{min}$ & $E_{max}$ & $z$ & $\Delta \gamma (E_{max},E_{min})$ \\
& & ms & MeV & MeV & & \\
\hline 
GRB 910711  &  BATSE         & 16  & 0.1  & 0.3  & $>$ 0.1 (assumed)  & $1.6 \times 10^{-13}$ \\%1.7E-13  \\
GRB 920229  &  BATSE         & 0.22 & 0.03 & 0.2  & $>$ 0.1 (assumed)  & $2.1 \times 10^{-15}$ \\%2.3E-15  \\
GRB 021206  &  RHESSI        & 4.8  &  3   & 10   & $>$ 0.15 (pseudo)       & $2.8 \times 10^{-14}$ \\%3.0E-14  \\
GRB 051221A &  Konus-Wind    & 4    & 0.07 & 0.3    & 0.547 (spectral)   & $4.7 \times 10^{-15}$ \\%4.9E-15   \\
GRB 090429  &  Swift         & 1200  & 0.25 & 0.5    & 6.5 (pseudo)       & $1.2 \times 10^{-13}$ \\ %7.1E-14   \\
GRB 090510  &  Fermi         & 1.0  & 1580 & 24,700 & 0.897 (spectral) & $6.6 \times 10^{-16}$ \\ %6.8E-16  \\   
\hline
\end{tabular}
\caption{A Table of data and WEP-violation limits for rapidly fluctuation and distant GRBs.}\label{tab:deltagammatable}
\end{center}
\end{table*}

These are the strongest limits yet found on WEP violations on high energy photons, and at cosmological distances. Further, the limit at $z > 6.5$ is the only limit when the universe was only a fraction of its present age. The strongest overall limit comes from GRB 090510 which is the strongest limit on WEP violation between any energy scale and anywhere in the universe.

In summary, by considering random galaxy clusters near the light path to the expected density, a distribution of gravitational potentials, and hence Shapiro time delays, along a random light path was simulated. We have shown that the contribution to the Shapiro time delay from these mass to any random light path is magnitudes greater than any local gravitational potential could provide, leading to an upper bound of $\Delta \gamma$ of WEP violations that is decreased significantly. This method can also be applied to future observations with potentially even finer time variability or across larger cosmological distances. Most importantly, this method is not unique to GRBs, but can be applied to any event that two particles are known to have been emitted, near simultaneously, across cosmological distances.

% \textit{Physical Review} style requires that the initial citation of
% figures or tables be in numerical order in text, so don't cite
% Fig.~\ref{fig:wide} until Fig.~\ref{fig:epsart} has been cited.

\begin{acknowledgments}
We thank Michigan Technological University for support during this research.
\end{acknowledgments}

\bibliography{WepPrl}
% Produces the bibliography via BibTeX.

\end{document}